\documentclass[dvips,11pt]{article}
\usepackage{graphicx,epsf,latexsym,makeidx}
\def\alga{Al$_x$Ga$_{1-x}$}

\def\sio2{$\mathrm{SiO_2}$}
\def\si29{$^{29}$Si}

\def\gq{$e^2/h~$}
\def\persec{$\mathrm{sec^{-1}}$}

\def\l2{{\leftarrow \atop \leftarrow}} 
\def\lr{{\leftarrow \atop \rightarrow}} 
\def\rl{{\rightarrow \atop \leftarrow}} 
\def\rr{{\rightarrow \atop \rightarrow}}

\def\uu{\uparrow \uparrow} 
\def\ud{\uparrow \downarrow} 
\def\du{\downarrow \uparrow}
\def\dd{\downarrow \downarrow}

\def\pt{(\lr +\rl) }
\def\ps{(\lr -\rl) }

\def\st{(\ud +\du) }
\def\ss{(\ud -\du) }

\def\11{| (\rr ) \ss \rangle }
\def\22{| \pt \ss \rangle }
\def\33{| \ps (\dd ) \rangle}
\def\44{| \ps \st \rangle}
\def\55{| \ps (\uu ) \rangle}

\begin{document}

\bf
\centerline{\large{Single Spin Measurement using Single Electron Transistors}}
\centerline{\large{to Probe Two Electron Systems}}
\rm
\medskip
\centerline{B. E. Kane*, N. S. McAlpine, A. S. Dzurak, R. G. Clark}
\medskip
\it
\centerline{Semiconductor Nanofabrication Facility}
\centerline{School of Physics, University of New South Wales}
\centerline{Sydney 2052 AUSTRALIA}
\medskip
\rm
\centerline{G. J. Milburn, He Bi Sun, and Howard Wiseman}
\medskip
\it
\centerline{School of Physics, University of Queensland}
\centerline{St. Lucia 4072 AUSTRALIA}
\medskip
\rm
\centerline{(\today)}

\medskip
\noindent
\begin{abstract}
We present a method for measuring single spins embedded in a solid by probing
two electron systems with a single electron transistor (SET).  Restrictions
imposed by the Pauli Principle on allowed two electron states mean that the
spin state of such systems has a profound impact on the orbital states
(positions) of the electrons, a parameter which SET's are extremely well
suited to measure.  We focus on a particular system capable of
being fabricated with current technology: a Te double
donor in Si adjacent to a Si/\sio2 interface and lying directly
beneath the SET island electrode, and we outline a
measurement strategy capable of resolving single electron and nuclear spins in
this system. 
We discuss the limitations of the measurement imposed by
spin scattering arising from fluctuations emanating from the SET
and from lattice phonons.  We conclude that measurement of
single spins, a necessary requirement for several proposed
quantum computer architectures, is feasible in Si using this strategy.

\medskip
\medskip

\noindent
PACS number(s): 85.30.Wx, 76.20.+q, 03.67.Lx

\medskip
\medskip

\noindent
*Current address: Laboratory for Physical Sciences, University
of Maryland, College Park MD 20740.  e-mail: kane@lps.umd.edu
\end{abstract} 

\section{Introduction}

The detection 
of a single electron or nuclear spin is perhaps the ultimate goal in
the development and refinement of sensitive measurement techniques in
solid state nanostructure devices.  
While of
interest in their own right, single spin measurements are particularly important
in the context of recently proposed solid state quantum computers, where
electron \cite{Loss98} \cite{Burkard99} \cite{Vrijen99}
and nuclear \cite{Kane98} spins are qubits which must
be initialized and measured in order to perform computation.  Methods proposed
for measuring single electron spins include using a sensitive magnetic
resonance atomic force microscope \cite{Sidles95} \cite{Wago97} and detecting
charge transfer across magnetic tunnel barriers \cite{Divincenzo98}.  Sensitive
optical techniques may also be promising \cite{Bonadeo98}.  Even if these techniques cannot
readily be integrated into a quantum computer architecture, single spin
measurements will be invaluable for measuring the electromagnetic environment
of the spin, which will determine the decoherence mechanisms ultimately
limiting a quantum computer's capability.

Here we discuss a method for probing the spin quantum numbers of a two electron system 
using a single electron transistor (SET).  Because of the Pauli Exclusion
Principle, spin quantum numbers of such systems profoundly affect the orbital
states (positions of the two electrons) of the system. Recently developed SET
devices are extraordinarily sensitive to charge configuration in the vicinity
of the SET island electrode, and they can consequently be used to measure the
spin state of two electron systems in appropriate circumstances.  In the scheme
previously proposed \cite{Kane98}, electron transfer into and out of bound
states on donors in Si are measured to determine whether the electrons are in a
relative singlet or triplet configuration.
SET's have already been proposed for performing quantum measurement on
qubits in a Josephson Junction-based quantum computer \cite{Shnirman98}.
SET's, operating at temperature $T \cong$100 mK,
have the recently demonstrated
capacity to measure charge to better than $10^{-4} e\mathrm{/\sqrt{Hz}}$ at
frequencies over 200 MHz \cite{Schoelkopf98}.  

Several material parameters make Si a good choice in which to fabricate single 
spin measuring devices: spin orbit coupling is small in Si, so the phonon-
induced spin lattice relaxation rate is almost seven orders of magnitude
smaller in Si \cite{Feher59b} than it is in GaAs \cite{Frenkel91}.  Also,
nuclear isotopes with nonzero spin can in principle be eliminated
in Si by isotope purification. 
The bound states on Si donors have been thoroughly characterized and studied. A
complication of Si arises from
its sixfold degenerate band structure.  We will focus on Si devices in this
paper, but the ideas presented here can be readily generalized to other
material systems. 

The configuration we  will
study is extremely simple (Fig. 1): a SET lies directly above two electrons
bound to a single donor impurity in an otherwise undoped 
layer of a Si
crystal.  Such a two-electron system, which can be thought of as a solid state
analog of a He atom, can be created in Si by doping with S, Se, Te, or Mg 
\cite{Grimmeiss86}\cite{Grossmann87}.  A \sio2 barrier layer isolates the SET
from the Si, and the substrate
is heavily doped, and hence conducting, beginning a few
hundred \AA~ below the donor. As drawn in Fig. 1 the device requires careful
alignment of the SET to the donor; however, the ideas in this paper could be
verified using a scanned probe SET \cite{Yoo97}, and the Te donor could be
deposited by ion implantation, so no nanofabrication on the Si would be
required.

The ground state of the electrons on the donor is a spin singlet.  The
experiment proceeds by applying a voltage between the SET and the substrate
just sufficient to ionize the donor and draw one electron towards the
interface. In this situation small changes in the applied voltage cause the
electron to move between the donor and the interface, and this electron motion
will change the SET conductance.  If the electrons are in a spin triplet state,
however, no bound state of appropriate energy exists on the donor, and
no charge motion will be observed. 
All the donors listed above have stable isotopes with
both zero and nonzero nuclear spin. If the
donor is a nucleus with nonzero spin, strong hyperfine
interactions couple the nuclear spin to the electrons,
and the nuclear spin can be inferred from measurements
of the motion of the electrons.
The measurement of both electron and nuclear spin  will require that the
electron Zeeman energy exceed $kT$ so that the electron spin states
are well resolved, a
condition which is readily met in Si at a temperature T$\approx$100 mK and
magnetic field $\mathbf{B}\approx$1 T. 

\section{Experimental Configuration}

Of the several possible two-electron donors in Si, we will focus on Te
for two reasons:  firstly, its 
energy levels are the shallowest of the Group VI donors
\cite{Grimmeiss86},
enabling it to be ionized by a relatively modest 
applied electric
field. 
Secondly, it is
a reasonably slow diffuser in Si \cite{Stumpel88}, and thus should be
compatible with most Si processing techniques.  The bound state energies
of Te donor states are shown in Fig. 2a.  $\mathrm{Te^0}$ and 
$\mathrm{Te^+}$ ground states are respectively 200 and 400 meV below
the conduction band.  

Electron orbital states in the Si conduction band have
a six-fold valley degeneracy, with valley minima located along the
[100] directions 85\% of the distance to the Brillouin zone boundary.  
This degeneracy is broken in states at a
donor by the central cell potential into a singly degenerate  $\mathrm{A_1}$
state, a triply degenerate  $\mathrm{T_2}$ state, and a doubly degenerate
E state.
The $\mathrm{A_1}$ state, which is a linear combination of each
of the six valleys, is the only state which has a finite probability
density
at the donor site, and
consequently has the lowest energy, owing to the central cell attractive
potential.  In $\mathrm{Te^0}$ two electrons lie in the $\mathrm{A_1}$
state in a
nondegenerate spin-singlet configuration.  This state is over 150 meV
below the excited states, including the lowest lying triplet configuration
of the two electron spins \cite{Grossmann87} \cite{Peale88}. 

In the proposed measurement configuration, an electric field $F$ is applied so
that an electron on the Te donor is weakly coupled to a state at a
[100] oriented
Si/\sio2 interface (Fig. 2b).   The condition that the donor and
interface states be weakly coupled requires that the distance
between the donor and the \sio2 interface must be 100-200 \AA.
Pulling the electron to the interface will thus require $F$=1-2
mV/\AA=0.1-0.2 MV/cm.  $F$ in the \sio2 layer will be approximately
three times bigger owing to the smaller dielectric constant in
\sio2. ($\epsilon_{\mathrm{Si}} \cong 12$; $\epsilon_{\mathrm{SiO_2}} \cong 4$.)
At these fields Fowler-Nordheim tunneling across a 100 \AA~ \sio2
barrier or between the Si valence and conduction band is
negligible \cite{Nagano94}, so charge will not leak into or out of the donor or
interface states.  The substrate must be $p$ doped, however, so that
the carriers in the substrate will be repelled from the interface
by $F$. 

When $F$=0, both electrons are bound to the Te donor (Fig. 3a);
however, one electron will occupy an interface state when $F$ is sufficiently
large (Fig. 3b).  In Si, the electron mass in each valley is anisotropic with
$m_\parallel=0.92~ m_0$ and  $m_\perp=0.19~ m_0$ \cite{Ando82}, masses
corresponding to motion parallel and perpendicular to the valley axis
respectively.   
At a [100] oriented Si/\sio2
interface, the sixfold valley degeneracy of electron states is broken,
and lowest energy states correspond to the two valleys along the axis
perpendicular to the interface.

When it is not located at the Te donor, the electron is still attracted
to the donor by its net positive charge.  While this attraction
is counteracted by $F$ in the $z$ direction, 
perpendicular to the interface, the electron
is drawn toward the donor in the $x-y$ plane,
resulting in the potential drawn in Fig. 3c. Thus, the electron
at the interface is still weakly bound to the donor.

The energies of the electron interface states will be the sum of
the binding energies in the $z$ and in the $x-y$ directions.
We assume that the $z$ confinement can be approximated by a
triangular potential.  The energies of the states are
\cite{Ando82}:
\begin{equation}
E_z (i)\cong \left\{ \frac{9 \pi^2}{8} \times \frac{\hbar^2 c^2}{m_z}
\times e^2 F^2 \times \left[ i- \frac{1}{4} \right]^2 \right\}^{\frac{1}{3}},
\end{equation}
for $i \ge$1.
For $m_z =m_\parallel$ and $F$=2 mV/\AA, $E_z (1)$=59 meV
and $E_z (2)$=104 meV.  The ground state
electron probability density function
is peaked at a distance $2 E _z (1)/ 3 eF \approx$ 20 \AA~
from the interface
and falls off rapidly at large distances.  The effect of the
donor a distance $z_0$=100-200 \AA~ from the interface is minimal on the
interface energy levels, but weak tunneling between the donor and the
interface is still possible.  For modeling of the system, we will
assume $z_0$=125 \AA.

The potential in the $x-y$ plane is:
\begin{equation}
U(r)=-\frac{e^2}{\epsilon_{eff.}} \times \left( r^2 + z^2_0 \right)^
{-\frac{1}{2}}.
 \end{equation}
Here, $r$ is the distance in the plane from the point in the plane
nearest the donor.  Because the electron sees an attractive
image charge associated with the Si/\sio2 dielectric boundary,
$\epsilon_{eff.}=(\epsilon_{\mathrm{SiO_2}}+
\epsilon_{\mathrm{Si}})/2=8$.  This potential is easily approximated
by a parabolic potential, leading to the following energies:
\begin{equation}
E_{xy} (j,k)= \frac{1}{2} \left( \frac{\hbar^2 e^2}
{\epsilon_{eff.} ~ m_{xy} ~ z^3_0 } \right)^{\frac{1}{2}}
\times (j+k),
 \end{equation}
for $j,k \ge 1$.
For $m_x=m_y=m_\perp$, $E_{xy} (1,1)$=6 meV and
$E_{xy} (1,2)$=9 meV.  The probability density for the parabolic
approximation wave function, plotted in Fig. 3c, is only large in the
region where the potential is well approximated by a parabola,
indicating that the parabolic approximation is justified.
An applied $\mathbf{B}\parallel z$ will modify these energies
significantly if the cyclotron energy,
$\hbar  \omega_c$, becomes comparable to the state
energy differences \cite{Ashoori96}.  However, at $\mathbf{B}$=1 T in Si,
$\hbar  \omega_c \approx$ 0.6 meV, so magnetic modification of
the orbital states should be minimal.  

These
results show that the lowest lying interface state is about 65 meV above the
conduction band, separated from the first excited state by $\approx$ 3meV.
These states are in the valleys along the $z$-axis.  Energies of 
the states in the valleys
along the $x$ and $y$ axes are $\sim$40 meV higher in energy. 
Because there are two valleys along the $z$ axis, the electron interface
states are still two-fold degenerate.  Sham and Nakayama \cite{Sham79}
have shown that this degeneracy is lifted by the sharp Si/\sio2
interface potential in the presence of an applied electric field.
They estimate $\Delta E_V \approx eF \times$ 0.5 \AA, corresponding to
a splitting of 1 meV for the proposed measurement configuration.
Although small, this splitting is sufficient to insure that the
interface electron occupies a single valley state at $T<$1K.

\section{Simplified Model Hamiltonian}

We model the
system using a simple Hamiltonian for the two electrons: they can be in only
two spatial states: either located at the donor $|\rightarrow\rangle$
or at the interface  $|\leftarrow\rangle$.  Additionally, the two electrons can
be in one of two spin states $|\uparrow\rangle$ or $|\downarrow\rangle$.  Of
the sixteen possible configuration states of two electrons in the
model, only six are
antisymmetric with respect to particle interchange, and are appropriate for
electrons.

Measurements will be made in the regime where the energy of the state in which 
both electrons lie on the donor, $|\rr \rangle$, is nearly degenerate with the
states in which one electron is at the donor and one is at the interface,
$|\rl \rangle$ and $|\lr \rangle$ .  The removal of both
electrons from the donor requires an additional 400 meV of energy (the
binding energy of the $\mathrm{Te^+}$
ground state).  
Consequently, we neglect the state  $|\l2 \rangle$  in which both electrons
are at the interface, since it is of much higher energy than the others.  The
five remaining antisymmetric basis states, eigenstates of both the particle
and spin exchange operator, are: 
\begin{equation}
\begin{array}{rcl}
 |1 \rangle & = &   \11 \\
 |2 \rangle & = &   \22  \\
 |3 \rangle & = &   \33  \\
 |4 \rangle & = &   \44 \\
 |5 \rangle & = &   \55, 
\end{array}
\end{equation}
where we have neglected normalization factors.  In the simplest
approximation, there are three terms in the Hamiltonian: $\Delta$, the
energy difference between the  $|\rr \rangle$ and the $|(\lr \pm \rl)\rangle$
states, can be varied by the bias applied between the substrate and
the SET island electrode.  The energy difference between $|\uparrow\rangle$
and $|\downarrow\rangle$ states is the Zeeman energy, $g \mu_B \mathbf{B}$
where $\mu_B$ is the Bohr magneton and $g$ is the Land{\'e} $g$ factor.
$t$ is the amplitude for the
electron to tunnel from the donor state to the interface state.  
The Hamiltonian matrix of the system is:
\begin{equation}
\label{H1}
\mathcal{H}_0=
\left( \begin{array}{ccccc}
\Delta & t & 0 & 0 & 0\\ 
t & 0 & 0 & 0 & 0\\ 
0 & 0 & -g \mu_B B & 0 & 0\\ 
0 & 0 & 0 & 0 & 0\\ 
0 & 0 & 0 & 0 & g \mu_B B
\end{array} \right).        
\end{equation}
The energy levels of this system,
plotted as a function of $\Delta$, are shown
in Fig. 4.  Because of the overall antisymmetry of the
electron wave function, the $|\rr \rangle$
state must be a spin singlet: $|\ss \rangle$.
A spin singlet state is also possible with a symmetric spatial state
of one electron on  the donor and
one at the interface $|\pt \rangle$.  
Hybridization of these
two levels results in the anticrossing behavior seen in Fig. 4.  In this system
the only possible spin triplet levels are associated with the spatially
antisymmetric state $|\ps \rangle$.  The energy of these three states,
although split by the magnetic field, are unaffected by the applied electric
field. Consequently, the spin singlet states are polarizable by an applied
electric field, while the spin triplet states are not.  This fact illustrates
how an electrical measurement can in principle determine a spin quantum
number.

\section{Measurement Procedure}
The difference in electric polarizability of singlet and triplet spin
states discussed above can be detected by a SET.~~SET's are
typically fabricated from Al,
with a small island electrode weakly coupled to two leads (the source and
drain) through thin $\mathrm{Al_2O_3}$ tunnel barrier layers (Fig. 1).  For
sufficiently small islands and at low temperatures the Coulomb blockade
prevents electron transport across the island unless a discrete energy level
of the island is resonant with the Fermi level in the source and drain. 
A SET can function as a sensitive electrometer because this resonance condition
is sensitive to any potentials coupling to the island - for example, coming
from the substrate in Fig. 1. The SET shown will exhibit periodic conductance
peaks with magnitude of order \gq as a function of substrate bias, each
corresponding to the addition of one electron to the island. Charge motion in
the vicinity of the SET changes the island potential and
results in shifts in the substrate bias voltage at which the peaks occur.

Figure 5 depicts both the energy levels of the two electron system as a
function of $\Delta$ and the conductance of the SET
as a function of substrate bias.  For simplicity we
assume that the SET conductance peaks are spaced symmetrically away from
the point where the electron levels cross ($\Delta$ = 0).  (The conductance
peaks can be moved to any position relative to the level crossing by
applying a voltage to an additional remote electrode, weakly coupled
capacitively to the SET.)  The
measurement proceeds by measuring the SET conductance on both sides of the
level crossing (at voltages $V_1$ and $V_2$)
by applying a voltage waveform to the
substrate similar to that shown in the inset to Fig. 5a. The measurement
must distinguish whether the electrons are in the 
lowest energy spin singlet or the lowest
energy spin triplet state.  At $V_2$  one electron is on the donor and one
electron is at the interface for both singlet and triplet states, so
the SET island potential - and hence the SET conductance - is the
same for both triplet and singlet states.  At $V_1$, however, the
singlet state is in a configuration where both electrons are on the
donor, while in the triplet state the electron positions are the same
as they were at $V_2$.  This difference in the electron positions
results in a difference in the potential at the island, and hence
a difference in the voltage at which the SET conductance maximum
occurs.  This conductance change can thus be used to infer the
spin state of the two electrons. 

The size of the offset between triplet and singlet conductance peak
positions is determined by how well the electrons are coupled to the
SET island and how far the electron moves. If the electron
moved all the way from the conducting substrate to the island, the
conductance peaks would be offset by one electron.  The 
approximate peak
position change for smaller electron movement is given by the
ratio $r$:
\begin{equation}
\label{r}
r=\left(\frac{z_0}{\epsilon_{\mathrm{si}}} \right) 
\times 
\left( \frac{w_{\mathrm{Si}}}{\epsilon_{\mathrm{si}}}+
\frac{w_{\mathrm{SiO_2}}}{\epsilon_{\mathrm{SiO_2}}} \right)^{-1},
\end{equation}
where $w_{\mathrm{SiO_2}}$ and $w_{\mathrm{Si}}$ are the thicknesses
of the \sio2 and undoped Si layers, respectively, and $z_0$ is the
distance which the electron moves.  For the layer thicknesses shown in Fig. 1
and $z_0$= 125 \AA, $r$=0.12.   Thus, the conductance peaks of the SET
can be offset approximately 10\% by the motion of the electrons
between the donor and interface states.

A charge sensitivity of 0.1 $e$ is readily achievable with SET's
and has been demonstrated with the recently developed RF-SET's
\cite{Schoelkopf98}, which are capable of fast ($>$100 MHz) measurements.
These RF-SET's have a demonstrated charge noise 
of $<5 \times 10^{-5} e\mathrm{/\sqrt{Hz}}$, implying that the
SET can measure 0.1$e$ in 0.25 $\mu$sec.
High speed operation of the
SET's may be necessary for the measurement because the
measurement must occur on a time short compared to the time the
electron scatters between spin states.  Spin scattering and
fluctuations are not included in the simplified Hamiltonian
of Eq. \ref{H1}
but will be present in real systems, and will be
discussed below.

In principle, a single conductance measurement at $V_1$
would be sufficient to determine the spin state of the
electrons, and the need to measure repeatedly at $V_1$
and at $V_2$ would be unnecessary.  However, motion of
remote charges will also couple to the 
SET \cite{Zimmerman97} leading to
drifting of the conductance peak positions (1/$f$
noise).  AC modulation of the substrate bias
can be used to measure the separation
between adjacent conductance peaks, rather than their absolute
position, and so can eliminate this
drift from the measurement.

\section{Effect of Fluctuations}

If the terms in Eq. \ref{H1} fluctuate, the energy levels shown
in Fig. 4 will not necessarily be eigenstates of the system, and transitions
between states will be possible.  Fluctuations will arise due to
lattice vibrations, and also will inevitably emanate from the SET, since
tunneling of electrons on and off of the SET island is a random process.
A rigorous approach to the effect of SET fluctuations must treat the
SET and the electrons being probed as a coupled quantum system.
Master equation techniques can be applied to this problem,
and have been used to analyze the system of a Josephson Junction
qubit coupled to a SET \cite{Shnirman98} and
tunneling devices \cite{Sun98}.  While a similar analysis of
a two electron system coupled to a SET is in preparation
\cite{Milburn99}, we will proceed by assuming that scattering of
the electrons is driven by external classical fluctuating fields,
the magnitudes of which we estimate from experimental conditions.
The scattering times so derived will then be compared to the
measurement time, derived above.

Fluctuations in  the occupancy of the SET island will
couple into the electron
system via $\Delta$, the potential difference between the donor and interface
states.  Phonon-induced fluctuations are, however,
the dominant mechanism of electron spin relaxation in lightly
doped Si, measured in electron spin resonance (ESR) experiments
\cite{Wilson61}.  The degeneracy of the six conduction band
valleys is broken by uniaxial stress directed along the [100]
directions, with compression lowering
the energy of the two valleys along the strain axis
with respect to the other four valleys.  To first order strain does not
affect the energy of the donor ground state, which is composed of equal amounts of each
of the six valleys, but the interface state energy level will shift with respect to the
donor state level with the application of strain. Thus, phonons will also lead to
fluctuations in $\Delta$.  Additionally, both bias and phonon fluctuations
couple to the $t$ term in the Hamiltonian, a mechanism of relaxation which
we will consider separately below. 
 
\section{Scattering between Spin Singlet States}

We treat the simplest case first, the effect of fluctuations in $\Delta$
on the two spin singlet states of Eq. \ref{H1}:
\begin{equation}
\label{H3}
\mathcal{H}=
\left( \begin{array}{cc}
\Delta & t \\ 
t & 0  
\end{array} \right).        
\end{equation}
The Hamiltonian is exactly diagonalized by rotating the
basis states though an angle $\chi=\tan^{-1}(2t/\Delta)$.
For fluctuations in $\Delta$, the relaxation rate between the
eigenstates of Eq. \ref{H3} is given by:
\begin{equation}
\label{g}
\Gamma = \frac{M^2}{4 \hbar^2} S_\Delta,
\end{equation}
where $M=\sin \chi$ and $S_\Delta$
is the spectral density of fluctuations of $\Delta$ evaluated at
the transition frequency between eigenstates.  The magnitude
of $M$ determines the degree to which the fluctuations couple between the
eigenstates, and scattering is reduced when $M \ll 1$.
Larger values of $|\Delta/t|$, far away from the anticrossing region,
will lead to smaller scattering rates between the
coupled singlet states if $S_\Delta$ is constant.

To determine an explicit value for $\Gamma$, we need to know $S_\Delta$.
For voltage noise emanating from the SET, $S_\Delta$ can be determined
from the time dependence of the charge on the SET island electrode.
The high frequency dynamics of SET's is still a topic of research,
and will depend sensitively on capacitances and inductances
of the SET
and in the external circuit.
To obtain crude estimates of relaxation times, we will simply assume
that SET noise is frequency independent shot noise determined
entirely by the SET current and the SET resistance:
\begin{equation}
S_V= S_I \times R^2=2 e I R^2 = 2 e V R,
\end{equation}
where $V$, $I$, and $R$ are the voltage, current and small signal resistance of the SET.
This leads to:
\begin{equation}
S_\Delta= 2 r^2 e^3 V R,
\end{equation}
where $r$, defined in Eq. \ref{r}, determines the proportion of
voltage that drops between the donor and interface states.

A quiescent SET, in which $V$=0, will generate a much smaller
amount of noise, especially if the island is biased so that
$R \rightarrow \infty$.  Again, for the purpose of generating
crude estimates, we assume that quiescent SET noise is given by
Johnson noise ($S_V=4 k T R$) when the SET is at a conductance peak.
To determine the magnitudes of shot noise and Johnson noise,
we use parameters tabulated by Schoelkopf for an optimized RF-SET
\cite{Schoelkopf98} biased to maximum sensitivity ($V \cong$ 1 mV,
$R$= 50 k$\Omega$, $T$=100 mK) in a configuration in
which $r$=0.1. Using these
numbers maximal scattering rates (using Eq. \ref{g} with $M$=1) are plotted in
Fig. 6.  Realistic values of the capacitance of the SET, which
has been entirely neglected in the foregoing, will tend to roll
off the spectra at frequencies $>$10 GHz.  Thus, the data
constitutes an upper bound on the scattering rates to be expected.

The magnitude of fluctuations in $\Delta$ induced by phonons is 
determined by $\Xi$, the deformation
potential, which is the rate the valley energy varies as strain
is applied, and by the density of states of phonons at a given
frequency.  A straightforward calculation leads to:
\begin{equation}
\label{phonon}
\Gamma_{phon.} = 
M^2 \times \nu^3 \coth \left( \frac{h \nu}{2 k T} \right)
\times \frac{8 \pi^3 \Xi^2}{h \rho} \left\{\frac{1}{v^5_l}+\frac{1}{v^5_t}
\right\},
\end{equation}
where $\rho$ is the density of the Si crystal and $v_l$ and $v_t$
are the velocities of longitudinal and transverse acoustic phonons
respectively.  Angular dependences of the phonon couplings have been
neglected in Eq. \ref{phonon}, as has the presence of the
nearby surface, which will modify the phonon spectrum
at low frequencies.
Thus, Eq. \ref{phonon} only
provides an approximate relaxation rate, which is plotted in Fig. 6 for
$T$=100 mK and $M$=1. 
This expression includes vacuum fluctuations, and is 
thus only appropriate
for transitions from higher to lower energy states when 
$h \nu > k T$.  While the phonon contribution to $\Delta$ rises
rapidly as a function of frequency, it only exceeds the shot noise
contribution at frequencies approaching 100 GHz.

To obtain approximate transition rates between the singlet states
using the shot noise expression,
we assume $\Delta/h$=100 GHz and $2 t/h$= 1 GHz, so $M^2 =10^{-4}$.
With these values, we obtain
$\Gamma = 10^{7}$ \persec~ or a decay time of 0.1 $\mu$sec.
This time is almost the same as the time estimated above for
RF-SET's to measure the spin state of the two electron
system.  It is likely that our use of a frequency independent
shot noise is an overestimate, and that the relaxation time
exceeds the measurement time.  Also, the measurement time
can possibly be reduced a factor of 10-100 in optimized SET devices
\cite{Schoelkopf98}.

\section{Scattering between Different Spin States}

At first glance, it would appear that the measurement time
$must$ be less than the singlet-singlet scattering time in
order for spin detection to be viable.  However, the point of
the measurement procedure is to distinguish between the
lowest lying singlet and triplet states.  Scattering between
these states (labeled ``3" in Fig. 5) must not occur.  However,
scattering between the other states can occur so long as the
average electron position difference between the singlet and triplet
states is resolvable.  Type 3 scattering must occur through
spin flips and in general will be much
weaker than scattering between the electric dipole coupled
singlet states. This is a crucial distinction between the
measurement of spin using SET's and the measurement of
charge quantum states, such as those in Josephson Junction
qubits \cite{Shnirman98},
where the states $to~be~distinguished$
are electric dipole coupled.

The Hamiltonian  of Eq. \ref{H1} is obviously oversimplified, since no terms
couple different spin states, and no spin relaxation is possible.
The dependence of the electron $g$ factor on external conditions, and
in particular on band structure parameters, is the major source of spin
relaxation in Si \cite{Wilson61} and consequently must be included in a
more accurate model of a two electron system in Si.  The extremely
long relaxation times measured in Si at low temperatures
($>$1000 sec.) \cite{Feher59b}
are a consequence of the fact these parameters are small in Si.
Additionally, if the electrons can exchange spin with other particles,
in particular with nuclear spins, then scattering between different
electron spin states will occur.  

The $g$ factor of an electron in a conduction band valley in Si  is not exactly
equal to the free electron value and is
slightly anisotropic, a consequences of spin orbit coupling.
The $g$ anisotropy
leads to a modified one electron spin Hamiltonian: 
\begin{equation}
H = \frac{1}{2} \mu_B B \left\{
g_{\parallel} \cos \theta ~ \mathbf{\sigma}_z +
g_{\perp} \sin \theta ~ \mathbf{\sigma}_x \right\},
\end{equation}
where $\mathbf{\sigma}$ are the Pauli spin matrices, and $\theta$
is the angle of $\mathbf{B}$ with respect to the valley ($z$) axis.
If the $z$ axis is redefined to be along $\mathbf{B}$, the
spin Hamiltonian becomes:
\begin{equation}
H = \frac{1}{2} \mu_B B \left\{
g_z \mathbf{\sigma}_z +
\beta \mathbf{\sigma}_x \right\},
\end{equation}
where:
\begin{equation}
g_z \equiv g_{\parallel} \cos^2 \theta + g_\perp \sin^2 \theta,
\end{equation}
and:
\begin{equation}
\beta \equiv (g_\perp - g_\parallel ) \sin \theta \cos \theta.
\end{equation}
The $g$ anisotropy will be the same for each of the two valleys
comprising the interface states; however, since the donor state
is an equal admixture of all six valleys, its $g$-factor will
be isotropic=$g_{0}$. For two electron systems, the spin dependent
corrections to the Hamiltonian in Eq. \ref{H1} are:
\begin{equation}
\label{H2}
{\mathcal{H}}'=\frac{1}{2}\mu_B B
\left( \begin{array}{ccccc}
0 & 0 & 0 & 0 & 0\\ 
0 & 0 & -\frac{\beta}{\sqrt{2}} & (g_{0}- g_z) & \frac{\beta}{\sqrt{2}}\\ 
0 & -\frac{\beta}{\sqrt{2}} & -(g_{0}+g_z) & \frac{\beta}{\sqrt{2}} & 0\\ 
0 & (g_{0} - g_z) & \frac{\beta}{\sqrt{2}} & 0 & \frac{\beta}{\sqrt{2}}\\ 
0 & \frac{\beta}{\sqrt{2}} & 0 & \frac{\beta}{\sqrt{2}} & (g_{0}+g_z)
\end{array} \right).        
\end{equation}
At the conduction band in Si, $g=(1/3)g_\parallel+(2/3)g_\perp$=1.99875
\cite{Feher59a}. $g_\parallel-g_\perp$, measured by applying strain to
shallow donors \cite{Wilson61}, is 1.0$\times10^{-3}$.  Finally,
$g_{0}$ for $\mathrm{Te}^+$=2.0023 \cite{Grimmeiss81}.  The
off-diagonal terms,
which will lead to scattering between spin
states if fluctuations are present,
are each $\cong10^{-3}$ and are small
perturbations on the original Hamiltonian.   The $\beta$ term will vanish,
in principle, if $\mathbf{B}$ is precisely aligned along a [100] axis of
the crystal, perpendicular to the interface, and this orientation
will presumably be the optimal experimental configuration.

To obtain an estimate for scattering rates between spin states,
an approximate solution to the full Hamiltonian (Eqs.
\ref{H1} and \ref{H2}) must be calculated.  The solution is
complicated by the fact that the $\22$ and $\44$ states
(the states with energy $\approx 0$ in Fig. 4)
are nearly degenerate when $t/\Delta \ll$1, a degeneracy
also weakly broken by the difference $(g_0-g_z)$
in $g$ factors at the
donor and interface states. To obtain an approximate
solution, valid in the measurement regime when  $t/\Delta \ll$1,
we first diagonalize the
Hamiltonian matrix to second order in $t$, which lowers the
$\22$ state energy with respect to  $\44$ by $t^2/\Delta$.
The $\22$ and $\44$ submatrix is then diagonalized exactly
by rotating the basis states through an angle $\xi$, where:
\begin{equation}
\tan \xi = (g_0-g_z)\mu_B B \times \frac{\Delta}{t^2}.
\end{equation}
Finally, corrections to the resultant wave functions are
determined to first order in the remaining $\beta$ terms
of Eq. \ref{H2}.

Once the perturbed wave functions are known, the matrix
elements coupling states generated by fluctuating terms
may be easily determined.  For a
fluctuation of the form $\Delta+\delta$ the perturbation
Hamiltonian matrix $\delta M_\Delta$ is given by:
\begin{equation}
\label{MD}
M_\Delta =
\left( \begin{array}{ccccc}
1 &
-(\frac{t}{\Delta}) \cos (\frac{\xi}{2}) &
-(\frac{\beta}{4\sqrt{2}})(\frac{t}{\Delta}) &
-(\frac{t}{\Delta}) \sin (\frac{\xi}{2}) &
-(\frac{\beta}{4\sqrt{2}})(\frac{t}{\Delta})\\
-(\frac{t}{\Delta}) \cos (\frac{\xi}{2})&
(\frac{t}{\Delta})^2 \cos^2 (\frac{\xi}{2}) &
(\frac{\beta}{4 \sqrt{2}})(\frac{t}{\Delta})^2 \cos (\frac{\xi}{2}) &
\frac{1}{2}(\frac{t}{\Delta})^2 \sin ( \xi ) &
(\frac{\beta}{4 \sqrt{2}})(\frac{t}{\Delta})^2 \cos (\frac{\xi}{2})\\  
-(\frac{\beta}{4 \sqrt{2}}) (\frac{t}{\Delta}) &
 (\frac{\beta}{4 \sqrt{2}})(\frac{t}{\Delta})^2 \cos (\frac{\xi}{2}) &
(\frac{\beta}{4 \sqrt{2}})^2 (\frac{t}{\Delta})^2 &
(\frac{\beta}{4 \sqrt{2}})(\frac{t}{\Delta})^2 \sin (\frac{\xi}{2}) &
(\frac{\beta}{4 \sqrt{2}})^2 (\frac{t}{\Delta})^2\\
 -(\frac{t}{\Delta}) \sin (\frac{\xi}{2}) &
\frac{1}{2}(\frac{t}{\Delta})^2 \sin ( \xi ) &
(\frac{\beta}{4 \sqrt{2}})(\frac{t}{\Delta})^2 \sin (\frac{\xi}{2}) &
(\frac{t}{\Delta})^2 \sin^2 (\frac{\xi}{2})  &
(\frac{\beta}{4 \sqrt{2}})(\frac{t}{\Delta})^2 \sin (\frac{\xi}{2})\\
-(\frac{\beta}{4 \sqrt{2}}) (\frac{t}{\Delta}) &
 (\frac{\beta}{4 \sqrt{2}})(\frac{t}{\Delta})^2 \cos (\frac{\xi}{2}) &
(\frac{\beta}{4 \sqrt{2}})^2 (\frac{t}{\Delta})^2 &
(\frac{\beta}{4 \sqrt{2}})(\frac{t}{\Delta})^2 \sin (\frac{\xi}{2}) &
(\frac{\beta}{4 \sqrt{2}})^2 (\frac{t}{\Delta})^2
\end{array} \right).        
\end{equation}
Only lowest order terms have been retained, and we have simplified
the expression by writing $(g_0+g_z)=4$. 
These matrix elements may be inserted directly into Eq. \ref{g}
to determine scattering rates between states induced by fluctuations
in $\Delta$. 

The matrix elements for scattering into and out of
$| \dd \rangle$ (state $| 3 \rangle$) contain a 
$\beta/(4 \sqrt{2})$ term in addition to the 
$t/\Delta$ present in the terms scattering
between the singlet states discussed above. Neglecting
entirely the angular dependence of $\beta$ and using
$g_\parallel-g_\perp =10^{-3}$, 
$(\beta/(4 \sqrt{2}))^2= 3 \times 10^{-8}$,
resulting in a
total scattering rate out of the
$| \dd \rangle$ state of about 0.3 \persec~ for the same
conditions used to calculate the scattering
rate between the singlet states above. 
This result suggests that very long averaging
times of the SET measurement will be possible before
spin relaxation occurs, and that single spin measurement
in Si will be possible in appropriately designed
devices.

In experimental conditions $\mathbf{B}$ will be sufficient to
effectively polarize the electrons, i.e. 
$g \mu_B B / k T \ge 10$. At $T$=100 mK, this requires that
$\mathbf{B}\cong$ 0.7 T and $\mu_B B \cong$10 GHz.
For $t/h$=1 GHz and $\Delta/h$=100 GHz, this implies that
$\tan \xi \cong$1 and that scattering to states
$\22$ and $\44$ (labeled respectively '1' and '2' in
Fig. 5) will be comparable.  As mentioned
above, this type of scattering will not harm the
measurement as long as the average positions
of electrons in the states being distinguished
differs.

\section{Scattering Induced by Fluctuations of $t$}

The calculation leading to Eq. \ref{MD} may be repeated
to determine the effect of fluctuations in $t$ on
scattering between states.  The result is:
\begin{equation}
M_t = \left( \begin{array}{ccccc}
(\frac{2 t}{\Delta}) &
\cos (\frac{\xi}{2}) &
-(\frac{\beta}{4\sqrt{2}}) &
\sin (\frac{\xi}{2}) &
-(\frac{\beta}{4\sqrt{2}})\\
\cos (\frac{\xi}{2})&
-(\frac{2 t}{\Delta})\cos^2 (\frac{\xi}{2}) &
-(\frac{\beta}{4 \sqrt{2}})(\frac{2 t}{\Delta}) \cos (\frac{\xi}{2}) &
-(\frac{t}{\Delta}) \sin ( \xi ) &
-(\frac{\beta}{4 \sqrt{2}})(\frac{2 t}{\Delta}) \cos (\frac{\xi}{2})\\  
(\frac{\beta}{4 \sqrt{2}})  &
-(\frac{\beta}{4 \sqrt{2}})(\frac{2 t}{\Delta})\cos (\frac{\xi}{2}) &
-(\frac{\beta}{4 \sqrt{2}})^2 (\frac{2 t}{\Delta})&
-(\frac{\beta}{4 \sqrt{2}})(\frac{2 t}{\Delta})\sin (\frac{\xi}{2}) &
-(\frac{\beta}{4 \sqrt{2}})^2 (\frac{2 t}{\Delta})\\
 \sin (\frac{\xi}{2}) &
-(\frac{t}{\Delta}) \sin ( \xi ) &
-(\frac{\beta}{4 \sqrt{2}})(\frac{2 t}{\Delta}) \sin (\frac{\xi}{2}) &
(\frac{2 t}{\Delta})\sin^2 (\frac{\xi}{2}) &
-(\frac{\beta}{4 \sqrt{2}})(\frac{2 t}{\Delta}) \sin (\frac{\xi}{2}) \\ 
(\frac{\beta}{4 \sqrt{2}})  &
-(\frac{\beta}{4 \sqrt{2}})(\frac{2 t}{\Delta})\cos (\frac{\xi}{2}) &
-(\frac{\beta}{4 \sqrt{2}})^2 (\frac{2 t}{\Delta})&
-(\frac{\beta}{4 \sqrt{2}})(\frac{2 t}{\Delta})\sin (\frac{\xi}{2}) &
-(\frac{\beta}{4 \sqrt{2}})^2 (\frac{2 t}{\Delta})
\end{array} \right).        
\end{equation}
Because of the absence of the $t/\Delta$ term present in
most of the matrix elements of Eq. \ref{MD}, fluctuations
in $t$ will have a greater effect on scattering than
fluctuations in $\Delta$ if the fluctuations are of
the same magnitude.

Band structure effects in Si can further magnify the importance
of $t$ fluctuations.  In Si the valley minima are located at
$k_0=0.85\times2 \pi/a$, where $a$=5.43 \AA~ is the lattice
constant.  If valleys on opposite sides of the Brillouin
zone are coupled to each other at two points in real
space (at two donors or at a donor and an interface),
standing waves with node spacing $\pi/k_0$ appear in
the coupling between the two sites.  These rapid
oscillations have been previously analyzed in the
context of the exchange interaction between donors
in doped Si \cite{Andres81}.  For a donor located
near an interface which breaks the valley degeneracy,
the coupling between the donor states and the two
valley states at the interface is a rapidly oscillating
function of the separation between the donor and the
interface (Fig 7).  If $t$ is a rapidly oscillating
function of external parameters, fluctuations in the
external parameters will be strongly amplified.

The magnitude of this effect may be most readily estimated
when the fluctuations arise from strain.  As mentioned
above, strain shifts the energies of the valleys along the
strain axis relative to the valleys on axes perpendicular
to the strain axis.  Strain, $s$, will
also change the value of $k_0$, the location of the valley
minima, and hence the wavelength of the standing waves.
We are unaware of measurements of $dk_0/ds$ but
estimate its order of magnitude by assuming that the effect of strain 
on electron energy levels is linear in $k_z$ in the neighborhood
of the valley minimum on the $z$ axis:
\begin{equation}
E(k_z)=\frac{\hbar^2}{2m_l}(k_z-k_0)^2+\frac{k_z}{k_0}\Xi s,
\end{equation}
where $z$ is the direction along the valley axis.
We entirely neglect effect of the orientation of the
applied strain.
Here, $\Xi$ is the deformation potential introduced above
= 9 eV in Si.  From this equation, we obtain:
\begin{equation}
\frac{dk_o}{ds}= - \frac{m_l \Xi}{\hbar^2 k_0}.
\end{equation}
Assuming $t=t_0\sin ( 2 k_0 z_0 )$, we obtain the maximum
effect of the strain as:
\begin{equation}
\left( \frac{dt}{ds} \right)_{max} = 
\left[ \frac{2 t_0 z_0 m_l}{\hbar^2 k_0} \right] \Xi.
\end{equation}
For $t_0/h$=1 GHz, and $z_0$=125 \AA, the term in 
brackets is $\cong 10^{-4}$.  This is the magnitude
of phonon-induced $t$ fluctuations relative to $\Delta$
fluctuations. For $t/\Delta = 10^{-2}$, the conditions
considered above, scattering rates attributable to fluctuations
in $\Delta$ will be four orders of magnitude larger than
those from $t$ fluctuations.  While the derivation leading
to this result is highly approximate, it does suggest that
$t$ fluctuations may be neglected, despite the amplifying
effect of oscillations induced by band structure.

Fluctuations in the voltage bias, or in the electric field
in the vicinity of the electrons, will also lead to fluctuations
in $t$.  It would seem that the effect of an electric field,
highly uniform on the scale of the lattice, would be small on
intervalley coupling.  However, the applied bias does
change the area on the interface where the electron wave function
is sizable, and the valley splitting induced by the interface will
be highly sensitive to the morphology of the interface, and hence
is very difficult to estimate. While we do not have a numerical
estimate for bias-induced $t$ fluctuations, it seems unlikely
that they will be an important source of scattering between
states. 

\section{Additional Sources of Noise in the
Electromagnetic Environment}

The major source of both electric dipole and
spin flip scattering in the electromagnetic
environment arises from the fluctuating
electric field generated by the SET. Because the
SET is a high impedance device, with resistance of
order $h/e^2$, the ratio of the magnetic to the electric
field generated by the SET is $\sim e^2/\hbar c$=1/137
in cgs units. $\mu_B B/(e z_0 F)$, the ratio of magnetic
to electric interaction energies of the SET with the
electrons is $\sim 10^{-7}$.  This leads to a
spin relaxation rate induced by the $magnetic$ field
emanating from the SET of $\sim 10^{-3}$ \persec,
which can be neglected.

A more relevant source of fluctuations arises because
RF-SET's are AC devices,
biased by a tuned circuit oscillating at $\nu \sim 1$
GHz.  GHz frequencies are employed to minimize the
contribution of noise from GaAs field effect
transistors (FET's) amplifying the
SET output.  Because the tuned circuit must be near
the SET, the device will be exposed to both
electric and magnetic fields at the SET bias frequency.
Consequently, the electron energy state differences will
need to be away from the SET bias frequency and its
harmonics during measurement.  Doing measurements at
magnetic fields when $g \mu_B B/h$=10-20 GHz should
fulfill this requirement.

\section{Scattering between States at Level Crossings}

To perform repeated measurements on the system, it will be
necessary to traverse the region where the two spin levels
cross one another (Figs. 4 and 5).    If there is a small
coupling between the two states, an anticrossing will occur,
and scattering will occur between the levels if the crossing
region is not traversed sufficiently rapidly. Ideally
however, the passage
should be ``adiabatic" with regard to the strongly coupled
singlet states, so that these states simply follow the levels
plotted in Figs. 4 and 5 as $\Delta$ is varied. These two
requirements imply that there is an optimal value for the
traversal rate, $\dot{\nu}$, where undesired scattering is
minimized.  As mentioned above, however, scattering between
the states being distinguished is much more harmful than
scattering between the singlet states, implying that
$\dot{\nu}$ be as large as possible.  Additionally,
even though the SET can be turned off during traversal,
some noise in the environment will be present during the
passage (the Johnson noise and the phonons, plotted in
Fig. 6), and a rapid traversal rate will minimize the
contribution of this noise to scattering at
the level crossing.

A simple Golden Rule calculation determines the scattering
probability $P$ between two crossing levels as a function  
of $\dot{\nu}$.  The result is:  
\begin{equation}
P=2\pi^2 \frac{\nu^2_{int}}{\dot{\nu}},
\end{equation}
where $h\nu_{int}$ is the energy difference between the
levels at the anticrossing point.  A likely upper limit
to the traversal rate is of order 100 GHz/nsec
=$10^{20} \mathrm{~Hz}^2$. We first estimate the
scattering resulting from the $\beta$ term in Eq.
\ref{H2}, again neglecting its angular dependence:  
$h\nu_{int} \cong \beta \mu_B B $=10 MHz.  These
values result in $P=2 \times 10^{-5}$.

Spin scattering can also occur near
the crossing point through the exchange of electron spin with
nuclear spins in the lattice, since natural Si contains
5\% \si29 with $I$=1/2.  The small value of
the nuclear Zeeman energy compared to the electron Zeeman
energy means that such scattering can only occur near
the level crossing point.
The electron interaction with \si29 will
be dominated by the contact hyperfine interaction
\cite{Slichter}:
\begin{equation}
\label{nu}
h\nu_A=\frac{8\pi}{3} \mu_B \mu_N g_N |\Psi(0)|^2,
\end{equation}
where $|\Psi(0)|^2$ is the electron probability density at the
nuclear site.  Evaluation of $P$ for  
the hyperfine interaction entails an appropriate average
over all lattice sites, assuming
that the total polarization of the nuclei is zero:
\begin{equation}
P=2\pi^2 \frac{\overline{\nu^2_{A}}}{\dot{\nu}}.
\end{equation}
The numerator in this expression 
is exactly the same average as that used to determine the mean
square line width of donor ESR lines, a parameter
which has been measured for Si:Te \cite{Grimmeiss81}.  In
$\mathrm{Te}^+$ using Si of natural isotopic composition,
the ESR line width is
$\sim$ 30 MHz,  leading to an estimate of $P\cong 10^{-4}$.
Interaction with lattice nuclei could be further reduced if necessary
by using Si depleted of \si29. These results imply that perhaps
thousands of passes across the level crossing can be made before
a spin scattering event occurs.

\section{Extension to Nuclear Spin Measurement}

In the foregoing discussion we have implicitly assumed that the nuclear spin
on the Te donor is zero.  While Te is composed of 92\% stable
$I=0$ isotopes, 7\% of natural Te is $^{125}\mathrm{Te}$, with $I$=1/2.
For the $\mathrm{Te}^+$ donor level, the electron spends approximately
10\% of its time on the donor site, and consequently 
$|\Psi(0)|^2$ in Eq. \ref{nu} can be large
\cite{Grimmeiss81}
\cite{Niklas83}.  For Si:$\mathrm{Te}^+$ the
zero $\mathbf{B}$ level splitting induced by hyperfine interactions is
3.6 GHz, which is comparable to the electron Zeeman splitting for
$\mathrm{B}$=0.1 T.  
	
The levels for a coupled two electron- and one nuclear-spin system are plotted
in  Fig. 8, with the small nuclear Zeeman energy splitting greatly exaggerated
so that the levels may be distinguished.  The electron Hamiltonian is that
of Eq. \ref{H1}, while the 
nucleus couples only to electrons at the donor site by the contact hyperfine
interaction.  The Hamiltonian again does not contain any terms that change the
total $z$ component of angular momentum of the system, and the state with all
spins pointing in the same direction (designated $|(\dd)(0) \rangle$
in Fig. 8) does not hybridize with other states.
The nuclear spin state in which the nuclear spin points
opposite to the electrons $|(\dd)(1) \rangle$ does hybridize with the electron
spin singlets that couple to the applied bias $\Delta$, leading to the
separation of the nuclear spin states shown in Fig. 8.

Measurement of the nuclear spin state proceeds in a manner entirely analogous 
to the spin measurement of the electron discussed above.  SET conductance
peak positions are measured at two fixed points on opposite sides of the level
crossing.  As in the case with electrons, scattering between the electric
dipole coupled states can occur during the measurement, so long as
scattering does not take place between the states being distinguished.
As in the case with electrons, these latter types of scattering processes
will occur as a result of electron $g$ fluctuations, impurity nuclear
spins, and nuclear and electron dipole interactions. Since the magnitudes
of these effects are similar for the electron and nuclear spin measurement
problem, 
it does not appear that measurement of nuclear spins will be intrinsically more
difficult than of electron spins.

\section{Experimental and Materials Issues}

We have focused on the Si/\sio2 material system for single spin measurement
devices, primarily because of the wealth of data in Si on ESR of donors.  These
ideas may be viable in other systems, and possibly in GaAs/\alga As
heterostructures, if the greater spin-orbit and hyperfine interactions in these
materials do not pose insurmountable problems. The lesser quality of the
Si/\sio2 interface compared to GaAs/\alga As should not affect the proposed
devices: a  mobility of $10^4 \mathrm{~cm^2/Vsec}$
implies energy fluctuations on the Si/\sio2 interface of order 0.5 meV,
less than the lateral binding energy of the interface electrons
to the donor calculated above.
We have neglected entirely the effects of the \sio2 layer on the
resonance and relaxation of the electrons.  ESR of conduction
electrons at the Si/\sio2 interface is very difficult to measure 
\cite{Stesmans93} \cite{Wallace91}, so experimental data on the
effect of the \sio2 interface is lacking.

Initial experiments will most simply be carried out on samples
randomly doped with Te by ion implantation or diffusion, and the
measurements made with a scanned SET so that many donors can
be tested for possible single spin sensitivity. Even if a
scanned probe SET is used, the material will have to be extraordinarily free
($\le 10^{10}/\mathrm{cm}^2$) of bulk and interface spin and
charge impurities in order to have a reasonable probability of
success in measuring a single spin, a requirement that may
prove very difficult to meet using conventional Si processing.
SiGe heterostructures may be an attractive alternative system \cite{Vrijen99}
if problems associated with interface states and dangling bonds
in Si/\sio2 structures prove to be insurmountable.

Finally, in order to demonstrate the measurement of a single
spin, the spin must first be prepared by placing it in a
known initial state.  For electrons, this can be accomplished
by simply waiting for a time long compared to the spin relaxation
time, so that the system will be in its lowest energy state with
high probability at low temperatures.
As shown in Fig. 5, the spin singlet is the
ground state at $V_1$ while the triplet is the ground state
at $V_2$, so the system can be prepared in either of the two
states by appropriately biasing the system and waiting a
sufficiently long time.

For nuclear spins, the relaxation times may be unreasonably long,
and the nuclear spin is best prepared by exposing the system
to an externally applied AC magnetic field $B_{AC}$
resonant with the
nuclear spin.  Action of  $B_{AC}$ can be used to flip the
nuclear spin from one state to another by appropriate pulses
or adiabatic passes across the resonance line prior to
the measurement process.  At higher frequencies an
applied $B_{AC}$ can also be used on the electrons,
and the small difference in the $g$ factor of the
donor and interface states allows particular electron
spins to be selectively flipped.

\section{Conclusion}

We have outlined a method for measuring single spin quantum numbers using 
single electron transistors in a Si solid state device that can be fabricated
with currently emerging technology.  While the impetus for realizing
these devices is the eventual development of a viable solid state quantum
computer technology, these devices will only be capable of very
rudimentary (single qubit, and perhaps two qubit) quantum logic.  They
should more appropriately be considered as solid state analogs of the
single ion traps, which have successfully demonstrated simple quantum logic
on single quantum states \cite{Monroe95}.  The analogy between these devices
and the single ion trap goes further, in that measurements are made in ion
traps by exciting transitions between the first of two
states being distinguished and a third state that is not coupled to the
second state.  If, and only if, the system is in the first state, many
``cycling transitions" are excited to the third state, allowing the states to
be distinguished with relative ease. In the devices discussed above, only one
of the two states being distinguished is electric dipole coupled to the
measuring SET, and the measurement process can continue until a
forbidden spin flip process occurs.

Also in analogy to the single
ion trap, these devices can be used to
measure the relaxation and decoherence processes operative on single
spins in solid state systems.  These measurements can be made by
using an applied $B_{AC}$ to perform $\pi$ and $\pi/2$
rotations on a single spin.  Such
measurements will be critical to determine whether quantum computation in
a solid state environment will be viable.  Aside from quantum computation,
precise measurement of single spins will be an extremely
sensitive probe of the electromagnetic environment of the spin, and may
have important heretofore unforeseen applications.

\newpage
\begin{figure}
\includegraphics[bb=0 0 595 770,width=12cm,clip=]{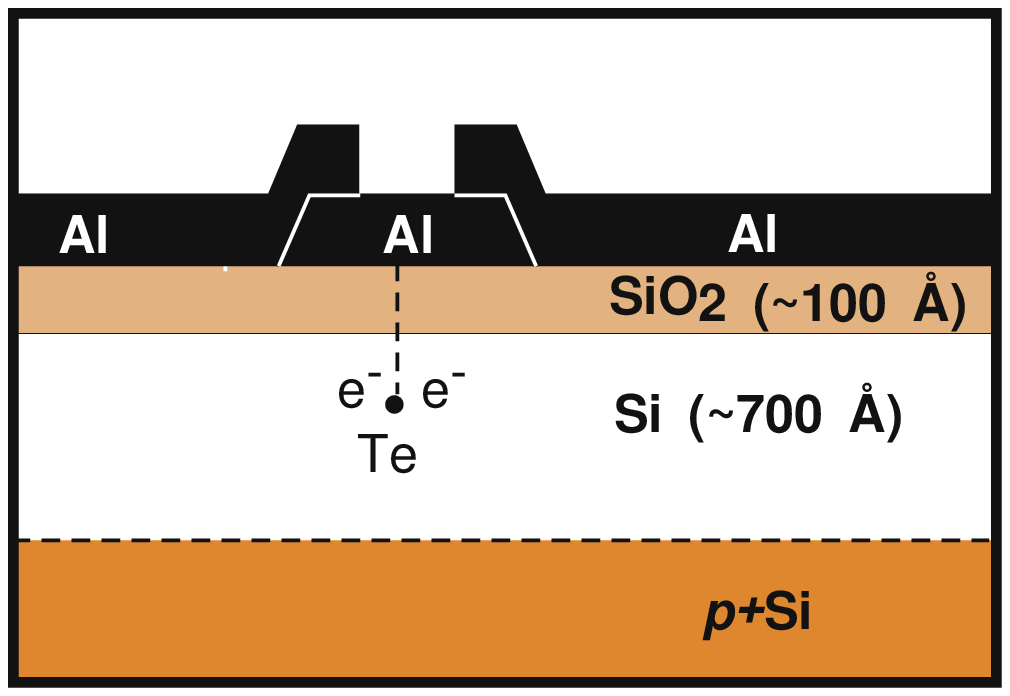}
\caption{Schematic diagram of the proposed measurement
configuration.
Conductance measurements are made on a single electron
transistor (SET), a device in which a small metallic island
electrode
(usually made of Al) is coupled to a source and drain
by tunnel junctions.  The
SET island lies directly 
above a Te double donor in Si, with a \sio2 barrier layer between the SET and
the donor.  A bias applied between the $p$-doped Si
substrate and the SET island
can pull one electron away from the donor into a state on the Si/\sio2
interface, a motion of charge which is detectable by the SET. In
demonstration experiments the SET could be at the tip of a scanned
probe, obviating the need to register the Te donor with the SET
island.}
\end{figure}

\newpage
\begin{figure}
\includegraphics[bb=0 0 595 870,width=12cm,clip=]{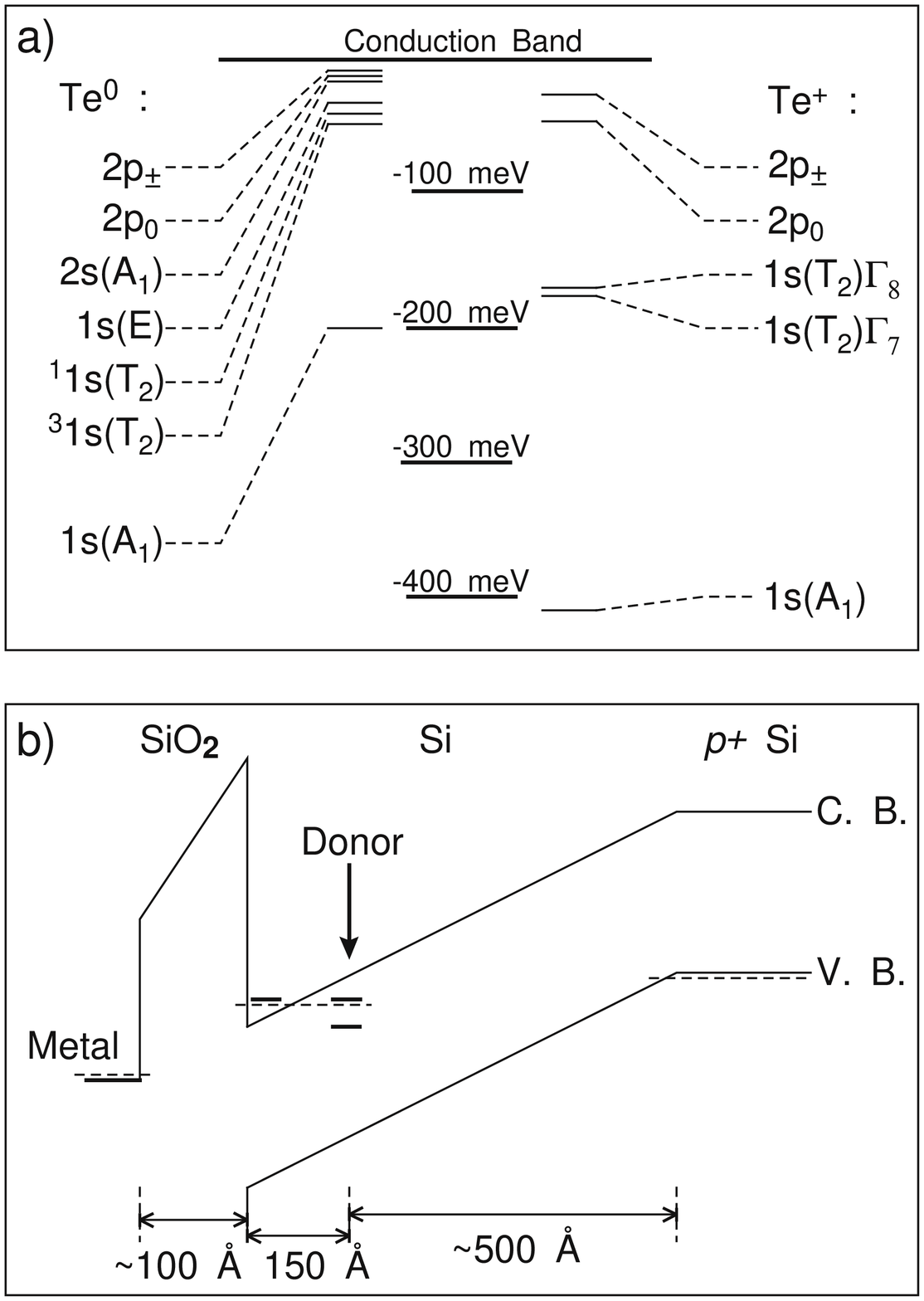}
\caption{(a) Energy levels of the neutral ($\mathrm{Te}^0$) and singly
ionized ($\mathrm{Te}^+$) states of a Te donor in Si.  The ground state
of $\mathrm{Te}^0$ is a spin singlet, 200 meV below the Si conduction
band.  Data is taken from Refs. \cite{Grossmann87} and \cite{Peale88}.
(b) Energy band diagram of the device.  An electric field $F$ is
applied between the Si substrate and the SET electrode sufficiently strong
to draw one electron away from the Te donor into a state on the 
Si/\sio2 interface.  The second electron remains bound to the donor.
The value of $F$ and the layer thicknesses specified ensure that
electron tunneling across the \sio2 interface and across the Si
band gap is negligible.  The substrate must be $p$ type, however, so that the
substrate carriers are not drawn towards the SET by the action of $F$. }
\end{figure}

\newpage
\begin{figure}
\includegraphics[bb=0 0 595 770,width=12cm,clip=]{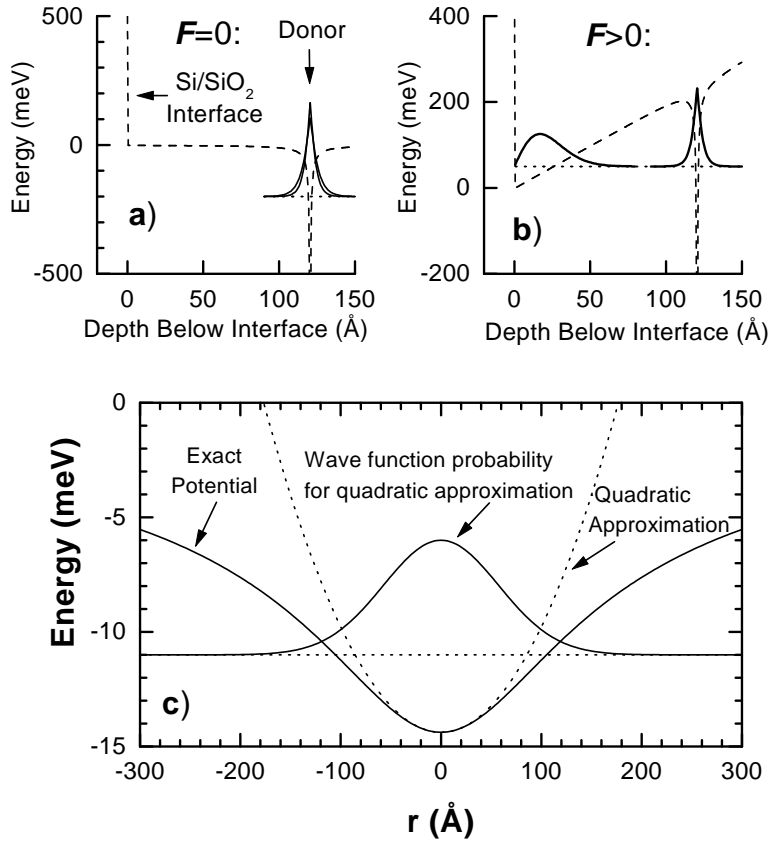}
\caption{(a-b)
Potential (dashed line) and electron wave functions
(solid lines)
depicted without and with an applied bias for a
donor at depth $z_0$= 125 \AA.  When
$F$=0 both electrons are bound to the donor.  At sufficiently large $F$, one
electron moves to a state at the interface, and has a wave function
characteristic of a triangular potential well.  (c) Potential and
electron probability in the $x-y$ plane at the interface.  When one
electron is at the interface, the donor has a net positive charge, so the
interface electron experiences an attractive potential in the x-y plane.
For the proposed devices the parabolic approximation
to the potential is reasonably valid. 
}
\end{figure}

\newpage
\begin{figure}
\includegraphics[bb=0 0 595 770,width=12cm,clip=]{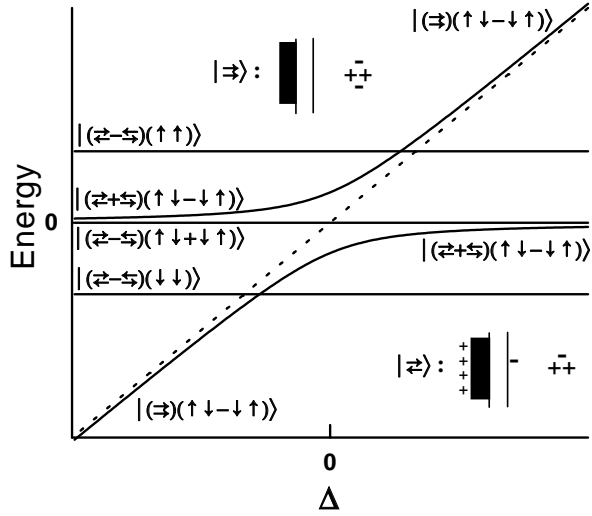}
\caption{Energy levels of the two electrons as a function of $\Delta$,
the energy difference between the two possible spatial configurations
of the electrons, using
the  simple Hamiltonian (Eq. \ref{H1})
discussed in the text.  Dotted line is the energy of the spin singlet
state in which both electrons are at the donor in the absence of
coupling between donor and interface states.  
When coupling is turned on, the two spin singlet states
hybridize, leading to anticrossing behavior seen in the graph.
The spin triplet states do not couple to the singlets, and are not
affected by $\Delta$, but are separated from each other by an external
magnetic field $\mathbf{B}$.}
\end{figure}

\newpage
\begin{figure}
\includegraphics[bb=0 0 595 770,width=12cm,clip=]{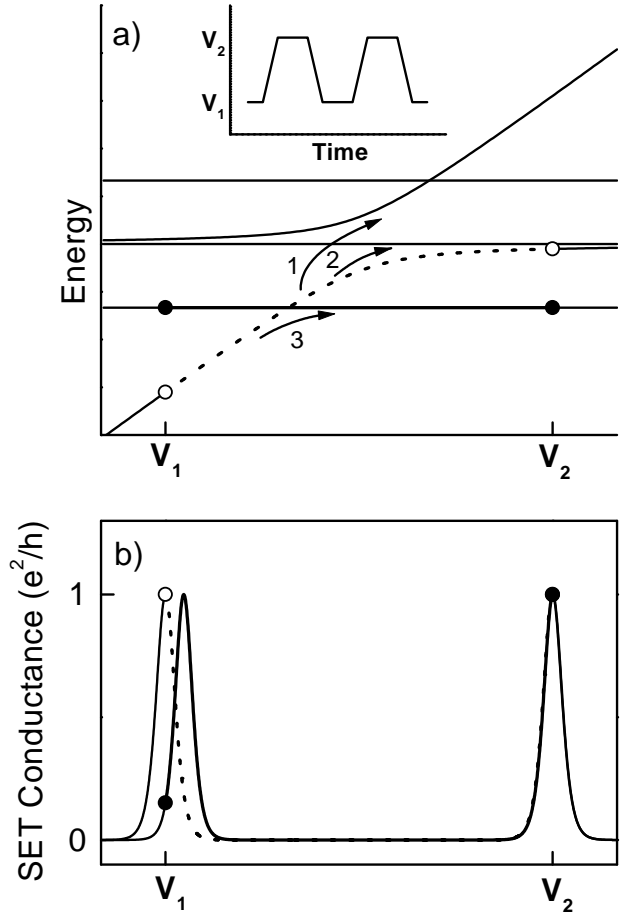}
\caption{To distinguish between the singlet and triplet spin
states, measurements are
made at two voltage biases, designated $V_1$ and $V_2$, well away from the
crossing points of the energy levels (a).  Charge position differs between
$V_1$ and $V_2$ for the singlet state (designated by a $\circ$), but not the
triplet state (designated by a $\bullet$).  (b) SET conductance as
a function of bias, showing characteristic peaks.  Positions of the
peaks are sensitive to the potential at the SET island.
The difference in charge position of the singlet and triplet states
results in a
difference in SET conductance at the measurement points.
This conductance difference can be measured repeatedly, using the
bias waveform shown in the top inset, to improve the signal to
noise ratio of the measurement.  Arrows in (a) designate
scattering mechanisms between states.  Type 3 scattering,
between the states being distinguished, must not occur before
the measurement process is completed.  Types 1 and 2
do not necessarily degrade the measurement, however.}
\end{figure}

\newpage
\begin{figure}
\includegraphics[bb=0 0 595 770,width=12cm,clip=]{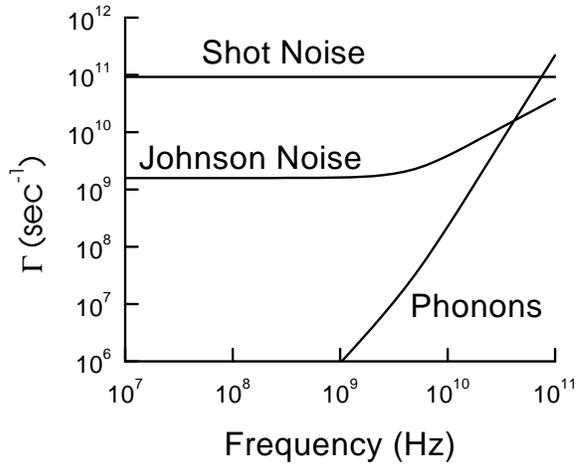}
\caption{Scattering rates attributable to shot noise, Johnson
noise and phonons.  The value for $\Gamma$ in Eq. \ref{g}
is plotted assuming $M$=1, i. e., coupling between states is
maximal.  Scattering arising from both shot noise and
Johnson noise will roll of at high frequencies as a result of
circuit capacitance, which has been neglected.}
\end{figure}

\newpage
\begin{figure}
\includegraphics[bb=0 0 595 770,width=12cm,clip=]{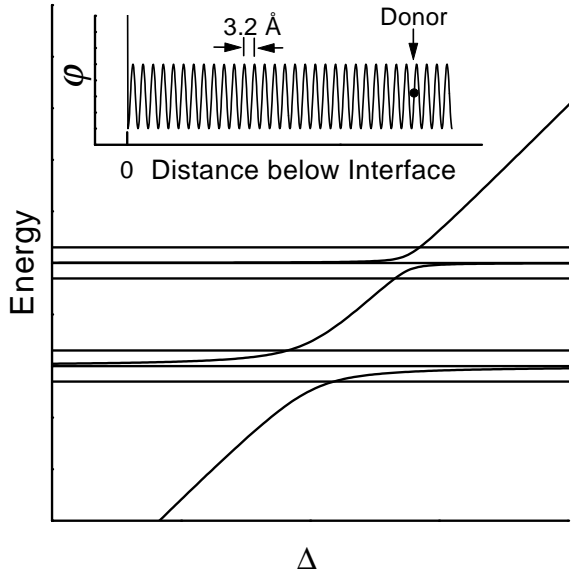}
\caption{Energy levels of the 
two-electron system explicitly showing the broken 
valley degeneracy of the interface states.  The
two valley electron states
are in fact a doublet, with approximate energy separation of 1 meV.  This
splitting is still large compared to the Zeeman energy, which creates the
triplet structure.  Because the valley phase of the donor states will differ
from the interface states in a fashion which varies rapidly with the
separation of the donor from the interface (inset),
the magnitude of the
coupling between the lowest energy interface state and the donor state is
oscillatory, and will be a rapidly varying periodic function of the
donor-interface separation.
}
\end{figure}

\newpage
\begin{figure}
\includegraphics[bb=0 0 595 770,width=12cm,clip=]{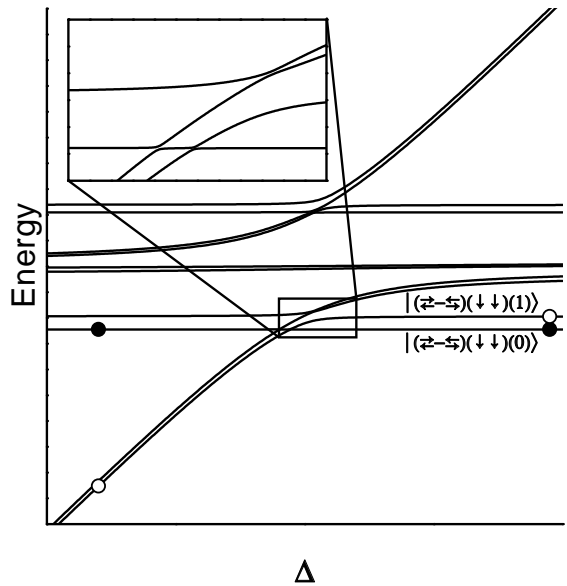}
\caption{Energy levels of a system of two electrons coupled by
the hyperfine interaction to a Te nucleus with spin 1/2.
Measurement of the
nuclear spin proceeds in a manner analogous to that for determining the
electron spin, with the charge configuration of the system measured by the SET
at two points on opposite sides of the level crossing.  Inset shows detail of
the level crossing region.}
\end{figure}

\end{document}